\newif\ifonecol

\onecoltrue    
\onecolfalse     

\ifonecol
\documentclass[11pt,draftcls,onecolumn,peerreview]{IEEEtran}
\else
\documentclass[journal]{IEEEtran}
\fi

\usepackage{amssymb,amsmath,xspace,intmacros,subfigure,amsfonts}
\usepackage[final]{graphicx}

\newcommand{\Bb}{{\textbf{B}}}
\newcommand{\Qb}{{\textbf{Q}}}
\newcommand{\xb}{{\textbf{x}}}
\newcommand{\sbb}{{\textbf{s}}}

\newcommand{\eb}{{\textbf{e}}}

\newcommand{\Gb}{{\textbf{G}}}
\newcommand{\Db}{{\textbf{D}}}
\newcommand{\Lb}{{\textbf{L}}}

\newcommand{\bb}{{\textbf{b}}}
\newcommand{\vb}{{\textbf{v}}}

\newcommand{\Hb}{{\textbf{H}}}
\newcommand{\Ib}{{\textbf{I}}}

\newcommand{\wb}{{\textbf{w}}}
\newcommand{\qb}{{\textbf{q}}}
\newcommand{\rb}{{\textbf{r}}}

\newcommand{\yb}{{\textbf{y}}}
\newcommand{\zb}{{\textbf{z}}}

\newcommand{\Phib}{{\mbox{\boldmath $\Phi$}}}
\newcommand{\phib}{{\mbox{\boldmath $\varphi$}}}

\newcommand\eg{e.g.,\xspace}
\newcommand\ie{i.e.,\xspace}

\newcommand{\lzero}{$\ell^{0}$}
\newcommand{\lone}{$\ell^{1}$}
\newcommand{\ltwo}{$\ell^{2}$}
\newcommand{\lp}{$\ell^{p}$}

\newcounter{examplecounter}
\renewcommand{\theexamplecounter}{\arabic{examplecounter}}

\newcommand{\inst}[1]{\unskip${^{#1}}$}

\begin{document}


\title{Bayesian Hypothesis Testing for Sparse Representation}
\author{H.~Zayyani \inst{1}, M.~Babaie-Zadeh\inst{1}~\IEEEmembership{Member} and C.~Jutten \inst{2}~\IEEEmembership{Fellow}\\
 \thanks{$^1$Electrical engineering department and Advanced Communication Research Institute (ACRI), Sharif university of
 technology, Tehran, Iran.}
 \thanks{$^2$GIPSA-lab, Grenoble, and Institut Universitaire de France, France.}
 \thanks{$^*$ This work has been partially funded by Iran NSF (INSF)
under contract number 86/994, by Iran Telecom Research Center
(ITRC), and also by center for International Research and
Collaboration (ISMO) and French embassy in Tehran in the framework
of a GundiShapour collaboration program.}
 }
\markboth{IEEE Trans Signal Processing, Vol. XX, No. Y, Month
2009}{Zayyani, Babaie-Zadeh and Jutten: Bayesian Hypothesis
testing}

\maketitle

\begin{abstract}
In this paper, we propose a Bayesian Hypothesis Testing Algorithm
(BHTA) for sparse representation. It uses the Bayesian framework
to determine active atoms in sparse representation of a signal.

The Bayesian hypothesis testing based on three assumptions,
determines the active atoms from the correlations and leads to the
activity measure as proposed in Iterative Detection Estimation
(IDE) algorithm. In fact, IDE uses an arbitrary decreasing
sequence of thresholds while the proposed algorithm is based on a
sequence which derived from hypothesis testing. So, Bayesian
hypothesis testing framework leads to an improved version of the
IDE algorithm.

The simulations show that Hard-version of our suggested algorithm
achieves one of the best results in terms of estimation accuracy
among the algorithms which have been implemented in our
simulations, while it has the greatest complexity in terms of
simulation time.

\end{abstract}

\section{Introduction}
\label{sec:intro} Finding (sufficiently) sparse solutions of
underdetermined systems of linear equations (possibly in the noisy
case) has been used extensively in signal processing community.
This problem has found applications in a wide range of diverse
fields. Some applications are Blind Source Separation (BSS) and
Sparse Component Analysis (SCA) \cite{ZibuP01}, \cite{GribL06},
decoding \cite{CandT05}, image de-noising \cite{EladA06}, sampling
and signal acquisition (compressed sensing) \cite{Dono06},
\cite{Bara07} and regression \cite{LarsS07}.

The problem can be stated in various contexts such as sparse
representation, SCA or Compressed Sensing (CS). Here, we use the
notation of sparse representation of signals. Let the model be:
\begin{equation}
\label{eq: model} \xb=\Phib\yb+\eb.
\end{equation}
where $\xb$ is an $n\times 1$ signal vector, $\yb$ is an $m\times
1$ sparse coefficient vector, $\Phib$ is an $n\times m$ matrix
called dictionary and $\eb$ is a $n\times 1$ error vector. It is
assumed that $n<m$ which means that the signal length is smaller
than the number of columns of the dictionary (which are called
atoms \cite{MallZ93}). So, the number of columns of the dictionary
is more than the number of rows of the dictionary, that is, the
dictionary is overcomplete. The main assumption is that the signal
has a sparse representation in this overcomplete dictionary. The
main goal is to find the sparse coefficient vector $\yb$ based on
the signal $\xb$ and knowing the dictionary $\Phib$. This problem
is nominated as sparse representation of the signal and the
methods are called sparse representation or sparse recovery
algorithms.

According to applications, the vector interpretations are
different, but in all of them the model follows (\ref{eq: model}).
For example, in the context of CS, $\Phib$ is the measurement
matrix, $\xb$ is a vector whose the few components are
measurements of the signal and $\yb$ is the sparse representation
of the true signal. In the context of SCA, $\Phib$ is the mixing
matrix, $\xb$ is the mixture vector and $\yb$ is the source
vector.

Because of $n<m$, there are usually infinitely many solutions of
this underdetermined system of linear equations. In the exact
sparse representation case, if we restrict ourselves to
sufficiently sparse coefficient vectors, it is proved that under
some conditions the sparsest solution is unique \cite{GoroR97},
\cite{DonoE03}, \cite{GribN03}. In the noisy case, there are
theoretical guarantees (in terms of sparse coefficients and
dictionary) for accurately and efficiently solving the problem
\cite{Trop06}, \cite{DonoET06}.

Finding the sparsest solution, that is, the solution with the
minimum number of nonzero elements, is an NP-hard combinatorial
problem. Different methods have been proposed to solve the problem
in a tractable way. Most of them can be divided in two main
categories: 1) Optimization approaches and 2) Greedy approaches
(or pursuit algorithms). The first category solves the problem by
optimizing a cost function according to different methods. The
second set of methods tries to find active coefficients (with
nonzero elements) directly through an algorithm.

The optimization approaches are basically split into convex and
non-convex optimization methods. The most successful approach
which is Basis Pursuit (BP) \cite{ChenDS98}, suggests a
convexification of the problem by replacing the
\lzero-norm\footnote{\lzero~norm of a vector is defined as the
number of its non-zero components. Although it is not a
mathematical norm, we use this name because it is frequently used
in the literature.} with the \lone-norm. It can then be
implemented by Linear Programming (LP) methods. Recently, a
Gradient-Projection algorithm for Sparse Reconstruction (GPSR) is
used for bound-constrained quadratic programming formulation of
these problems \cite{FiguNW07}. A method for large scale
\lone-Regularized Least Square (\lone-RLS) is also devised in
\cite{KimKLBG07}. In addition, an Iterative Bayesian Algorithm
(IBA) is used for solving the problem with a convex cost function
which its steps resemble E-step and M-step of an EM algorithm
\cite{ZayyBJ08},~\cite{ZayyBJ08ICASSP}.

Among the nonconvex cost function methods, the FOCUSS algorithm
uses \lp-norm with $p\leq1$ instead of \lzero-norm in the
noise-free case~\cite{GoroR97}, \cite{RaoD99}. Regularized-FOCUSS
(R-FOCUSS) method extends FOCUSS for the noisy case with a
Bayesian framework \cite{RaoECPD03}. There are also some Bayesian
methods such as Relevance Vector Machine (RVM) \cite{Tipp01},
Sparse Bayesian Learning (SBL) \cite{WipfR04} and recently a
Bayesian Compressive Sensing approach (BCS) \cite{JiXC08} which
mainly solve a nonconvex problem. Recently, a smoothed version of
the \lzero-norm was used for solving the problem by a
gradient-ascent method which is called Smoothed-\lzero (SL0)
\cite{MohiBJ08}. Moreover, a Sparse Reconstruction by Separable
Approximation (SpaRSA) algorithm is suggested for group separable
regularizers which is usually nonsmooth and possibly also
nonconvex \cite{WrigNF08}. There is also an Iterative Reweighted
Algorithm for nonconvex CS (IRA) \cite{CharY08}.

The other category is the greedy algorithms which choose
successively the active coefficients without having any explicit
cost function. Generally, they use the correlation between the
signal (or residual signal) and the atoms of the dictionary as an
informative measure for deciding which coefficients are actually
active (or nonzero). These algorithms are Matching Pursuit (MP)
\cite{MallZ93}, Orthogonal Matching Pursuit (OMP) \cite{PatiRK93},
Stage-wise OMP (StOMP) \cite{DonoTDS06}, Weighted MP (WMP)
\cite{EscoGV06}, Tree-Based Pursuit (TBP) \cite{JostVF06},
Regularized OMP (ROMP) \cite{NeedV07}, Gradient Pursuit (GP)
\cite{BlumD081}, Stagewise weak Gradient Pursuit (StGP)
\cite{BlumD082} and Compressive Sampling MP (CoSaMP)
\cite{NeedT08}.

Besides these two main approaches, one can mention Iterative
THresholding algorithms (ITH) \cite{DaubDD04}, an Iterative
Detection Estimation (IDE) method \cite{AminBJ06} and three
Minimum Mean Square Estimation (MMSE) algorithms \cite{LarsS07},
\cite{SchnPZ08} which can be considered as Bayesian approaches
which use discrete search techniques for finding the dominant
posteriors. In \cite{LarsS07}, an algorithm is proposed for
Approximating the MMSE estimate (A-MMSE) for the sparse vector in
the application of linear regression. \cite{SchnPZ08} also
presents a Fast Bayesian Matching Pursuit (FBMP) method for
recursive MMSE estimation in linear regression models.

Table~\ref{table1} shows the overall sparse representation
algorithms that we have mentioned. In this table, Bayesian methods
are highlighted with bold characters. One can consider the
Bayesian methods as a distinct category, but we did not do that
because they also need some kind of cost function and optimization
techniques or algorithms to solve their problem.

\begin{table}[tb]
\centering%
\caption{%
    Sparse Representation algorithms ( bold names are Bayesian approaches).%
}
\begin{tabular}{c | c | c | c}
\hline \hline
        \multicolumn{2}{c}{Optimization algorithms} & \multicolumn{1}{c}{Greedy} & Other\\
        \multicolumn{1}{c}{Convex} & \multicolumn{1}{c}{Nonconvex} &\multicolumn{1}{c}{Algorithms} & Algorithms \\
\hline  BP \cite{ChenDS98}     & FOCUSS \cite{GoroR97}, \cite{RaoD99}      & MP \cite{MallZ93} &  ITH \cite{DaubDD04}    \\
        GPSR \cite{FiguNW07}   & \textbf{R-FOCUSS} \cite{RaoECPD03}      & OMP \cite{PatiRK93} &  IDE \cite{AminBJ06}     \\
        \lone-RLS \cite{KimKLBG07}   & \textbf{RVM} \cite{Tipp01}      & StOMP \cite{DonoTDS06} & \textbf{A-MMSE} \cite{LarsS07}      \\
        \textbf{IBA} \cite{ZayyBJ08} & \textbf{SBL} \cite{WipfR04}      & \textbf{WMP} \cite{EscoGV06} &  \textbf{FBMP} \cite{SchnPZ08}  \\
        &  \textbf{BCS} \cite{JiXC08}     & TBP \cite{JostVF06} &    \\
          &  SL0 \cite{MohiBJ08}     & ROMP \cite{NeedV07} &     \\
          & SpaRSA \cite{WrigNF08}      & GP \cite{BlumD081} &     \\
         & IRA \cite{CharY08}      & StGP \cite{BlumD082} &     \\
          &       & CoSaMP \cite{NeedT08} &     \\
\end{tabular}
\label{table1}
\end{table}

An important task in the sparse representation is to determine
which coefficients are nonzero or in other words which atoms are
active in the sparse representation of the signal. This is mainly
done with Correlation Maximization (CM) in the pursuit algorithms
with some differences. So, the core idea of the pursuit algorithms
is to use the correlation of the residual signal with the atoms to
determine the active atoms. For example, MP uses the CM to select
at each iteration one active atom. StOMP uses a thresholding to
select several active atoms at a same time. Other methods like
IDE, use a measure of activity to determine the corresponding
nonzero coefficients. The IBA algorithm \cite{ZayyBJ08} uses a
steepest-ascent to determine a vector which is defined as the
activity vector.

The simplicity of the greedy algorithms or pursuit algorithms
arises in determining one active atom (\eg in MP) or several
active atoms (\eg in StOMP) at an instant. So, they determine the
activity vector in a simple way rather than to solve a hard
optimization problem in a multi-dimensional space. The basic idea
of this paper is to use the correlation between the signal and
atoms like pursuit algorithms. Then, a Bayesian hypothesis test is
used to estimate the activity measure for each coefficient
separately. So, the aim of this paper is to estimate simple
activity measures using a Bayesian framework. This is done by
three simple assumptions which are needed when we devise our
algorithm. These assumptions are just approximations and the
algorithm is devised under these simplifying assumptions. The
results of this work have been partially presented in
\cite{ZayyBJ09}.

The activity measure we obtain in this method is similar to what
has already been obtained by IDE algorithm \cite{AminBJ06}. The
main difference, however, is that the threshold is obtained
mathematically and is calculated throughout the algorithm by some
simple parameter estimation techniques.

In this paper, we first introduce our system model and some
notations in Section~\ref{sec: sysmodel}. Then, in
Section~\ref{sec: BPA}, we propose our Bayesian Hypothesis Testing
Algorithm (BHTA). Section~\ref{sec: conv} investigates the
stability analysis of the algorithm. Finally, in Section~\ref{sec:
simresult}, we investigate the experimental performance of the
BHTA in comparison with other main algorithms.

\section{System model}
\label{sec: sysmodel} The noise vector $\eb$ in (\ref{eq: model})
is assumed to be zero-mean Gaussian with covariance matrix
$\sigma_e^{2}\mathbf{I}$. In the model, the coefficients are
inactive with probability $p$, and are active with probability
$1-p$ (sparsity of $\yb$ implies that $p$ should be near 1). In
the inactive case, the values of the coefficients are zero and in
the active case the values are obtained from a Gaussian
distribution. We call this model the `spiky model' which is a
special case of the Bernoulli-Gaussian model with the variance of
the inactive samples being zero. This model has been also used
in~\cite{GeorM97} and~\cite{LarsS07}. It is suitable for sparse
representation of a signal where we would like to decompose a
signal as a combination of only a few atoms of the dictionary and
the coefficients of the other atoms are zero. So, the probability
density of the coefficients in our problem is:
\begin{equation}
\label{eq: spiky} p(y_i)=p\delta(y_i)+(1-p)N(0,\sigma_r^{2}).
\end{equation}
where $\delta(.)$ denotes the Dirac impulse function. In this
model, each coefficient can be written as $y_i=q_ir_i$ where $q_i$
is a binary variable (with a binomial distribution) and $r_i$ is
the amplitude of the $i$'th coefficient with a Gaussian
distribution. Each element $q_i$ is the activity of the
corresponding coefficient (or corresponding atom):
\begin{equation}
\label{eq: activity}
   q_i=\left\{\begin{array}{ll}
   1 & \textrm{if $y_i$ is active (with probability $1-p$)}\\
   0 & \textrm{if $y_i$ is inactive (with probability $p$)}
   \end{array}. \right.
\end{equation}
Consequently, the probability $p(\mathbf{q})$ of the activity
vector $\qb\triangleq(q_1,q_2,...,q_m)^T$ is equal to:
\begin{equation}
\label{eq: pq} p(\mathbf{q})=(1-p)^{n_a}(p)^{m-n_a}.
\end{equation}
where $n_a$ is the number of active coefficients, \ie the number
of 1's in $\mathbf{q}$.  So, the coefficient vector can be written
as:
\begin{equation}
\label{eq: SQR relation} \mathbf{y}=\mathbf{Q}\mathbf{r}.
\end{equation}
where $\mathbf{Q}=\diag(\mathbf{q})$ and
$\rb\triangleq(r_1,r_2,...,r_m)^T$ is the `amplitude vector'. Note
that, in this paper, we use the same notation $p(\cdot)$ for both
probability and Probability Density Function (PDF).

\section{Bayesian Hypothesis Testing Algorithm (BHTA)}
\label{sec: BPA} 
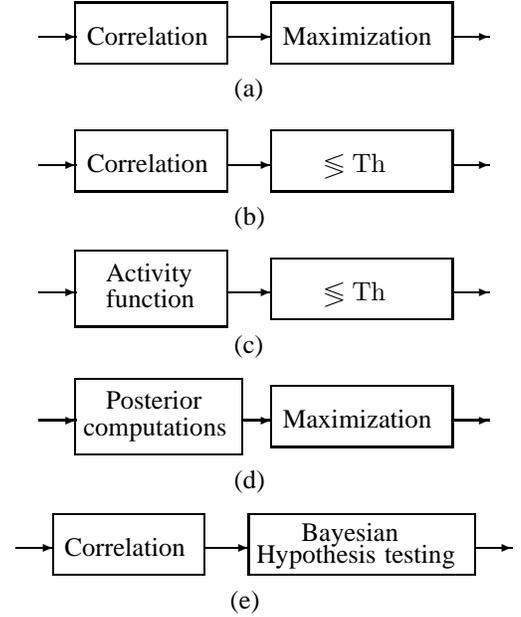
\begin{figure}[tb]
\flushleft \unitlength 0.1cm \linethickness{0.4pt}
\begin{center}

\begin{picture}(24,17)(0,-0.2)
 \put(-9,4.5){\framebox(20,8)[cc]}
 \put(-7.5,7){Correlation}

 \put(17,4.5){\framebox(24,8)[cc]}
 \put(18.5,7){Maximization}
 \put(12,0){(a)}

 \put(-14,8){\vector(1,0){5}}
 \put(11,8){\vector(1,0){6}}
 \put(41,8){\vector(1,0){5}}

\end{picture}

\begin{picture}(24,17)(0,-0.2)
 \put(-9,4.5){\framebox(20,8)[cc]}
 \put(-7.5,7){Correlation}

 \put(17,4.5){\framebox(24,8)[cc]}
 \put(24,7){$\lessgtr \mathrm{Th}$}
 \put(12,0){(b)}

 \put(-14,8){\vector(1,0){5}}
 \put(11,8){\vector(1,0){6}}
 \put(41,8){\vector(1,0){5}}

\end{picture}

\begin{picture}(24,17)(0,-0.2)
 \put(-9,3.5){\framebox(20,10)[cc]}
 \put(-5,9.5){Activity}
 \put(-5,6){function}

 \put(17,4.5){\framebox(24,8)[cc]}
 \put(24,7){$\lessgtr \mathrm{Th}$}
 \put(12,0){(c)}

 \put(-14,8){\vector(1,0){5}}
 \put(11,8){\vector(1,0){6}}
 \put(41,8){\vector(1,0){5}}

\end{picture}

\begin{picture}(24,17)(0,-0.2)
 \put(-9,3.5){\framebox(22,10)[cc]}
 \put(-5,9.5){Posterior}
 \put(-8,6){computations}

 \put(17,4.5){\framebox(24,8)[cc]}
 \put(18.5,7){Maximization}
 \put(12,-1){(d)}

 \put(-14,8){\vector(1,0){5}}
 \put(13,8){\vector(1,0){4}}
 \put(41,8){\vector(1,0){5}}

\end{picture}

\begin{picture}(18,17)(0,-0.2)
 \put(-15,4.5){\framebox(20,8)[cc]}
 \put(-13.5,7){Correlation}

 \put(11,4.5){\framebox(30,8)[cc]}
 \put(18,9){Bayesian}
 \put(12,6){Hypothesis testing}
 \put(8.5,0){(e)}

 \put(-20,8){\vector(1,0){5}}
 \put(5,8){\vector(1,0){6}}
 \put(41,8){\vector(1,0){5}}

\end{picture}

\end{center}
\caption{\footnotesize Idea of detection in various algorithms.
(a) MP or OMP (b) StOMP (c) IDE (d) MAP sense (\eg IBA) (e)
Bayesian Hypothesis Testing Algorithm (BHTA).} \label{fig1}
\end{figure}

The main task in sparse representation algorithms
is to determine which atoms are active in the sparse
representation of the signal. This can be viewed as a detection
task like in the IDE algorithm \cite{AminBJ06} which an activity
function is compared with a decreasing threshold. In some pursuit
algorithms (\eg MP), it is determined by Correlation Maximization
(CM). In some other pursuit algorithms (\eg StOMP), it is done by
comparing the correlations with a threshold. In the MAP sense, it
is done with posterior maximization over all possible activity
vectors \cite{ZayyBJ07DSP}. In IBA algorithm \cite{ZayyBJ08}, the
maximization is done by a steepest-ascent algorithm in the M-step
within a MAP sense framework. Here we want to determine the
activity by a Bayesian hypothesis testing from the correlations.
The possible strategies for determining the active atoms for the
various algorithms are schematically depicted in
Fig.~\ref{fig1}(a)-(e).

To develop a hypothesis testing approach, we write (\ref{eq:
model}) as:
\begin{equation}
\xb=\sum_{i=1}^m\phib_iy_i+\eb.
\end{equation}
where $\boldsymbol{\phi}_i$ is the $i$'th column (\ie the $i$'th
atom) of the dictionary. So, the correlations between the original
signal and the atoms are:
\begin{equation}
\label{eq: BP1}
z_j\triangleq<\xb,\boldsymbol{\phi}_j>=y_j+\sum_{\substack{i=1\\i\neq j}}^my_ib_{ij}+v_j.
\end{equation}
where $b_{ij}\triangleq<\boldsymbol{\phi}_i,\boldsymbol{\phi}_j>$
and $v_j\triangleq <\eb,\boldsymbol{\phi}_j>$, and the atoms are
assumed to have unit Euclidean norm.

To do a Bayesian hypothesis test based on correlations for
determining the activity of the $j$'th atom, we must compute the
posteriors $p(H_1|\zb)$ and $p(H_2|\zb)$, where $H_1$ is the
hypothesis that the $j$'th atom is active and $H_2$ is the
hypothesis that the $j$'th atom is inactive. To obtain a simple
algorithm like pursuit algorithms, assuming the previous
estimations of all other coefficients (except the $j$'th
coefficient), we want to detect the activity of only the $j$'th
atom and then update only the $j$'th coefficient.

Since we assume that we know previous estimations of other
coefficients, (\ref{eq: BP1}) can be written as:
\begin{equation}
z_j-\sum_{i=1, i\neq j}^m\hat{y}_ib_{ij}=y_j+\sum_{i=1, i\neq
j}^m(y_i-\hat{y}_i)b_{ij}+v_j.
\end{equation}
where $\hat{y}_i$ is the estimation of the $i$'th coefficient at
the current iteration. Let define:
\begin{equation}
c_j\triangleq \sum_{i=1, i\neq j}^m\hat{y}_ib_{ij}.
\end{equation}
\begin{equation}
\label{eq: gamma} \gamma_j\triangleq \sum_{i=1, i\neq
j}^m(y_i-\hat{y}_i)b_{ij}+v_j.
\end{equation}
The two hypotheses $H_1$ and $H_2$ are then:
\begin{equation}
\label{eq: hyp}
\text{Hypotheses}:\left\{\begin{array}{ll}
     H_1: z_j-c_j=r_j+\gamma_j \\
     H_2: z_j-c_j=\gamma_j
     \end{array}. \right.
\end{equation}
where $c_j$ is known and $\gamma_j$ is a noise or error term. In
fact, Eq.~(\ref{eq: hyp}) is a classical detection problem.

\subsection{Hard-BHTA}
\label{sec: HBPA} In this section, we suggest a classical
detection solution for solving the problem (\ref{eq: hyp}). As it
was said before, the hypothesis test involves the computation of
the overall posteriors $p(H_1|\zb)$ and $p(H_2|\zb)$. But, with
the previous formulations, we reach a relatively simple detection
problem as in (\ref{eq: hyp}). For the simplicity of the algorithm
like the pursuit algorithms, we rely only on the correspondent
correlation (\eg $z_j$) and hence the simpler posteriors as
$p(H_1|z_j)$ and $p(H_2|z_j)$. So, the hypothesis $H_1$ is chosen
when $p(H_1|z_j)>p(H_2|z_j)$, otherwise $H_2$ is chosen.

Based on Bayes' rule, the above posteriors are proportional to
$p(H_1)p(z_j|H_1)$ and $p(H_2)p(z_j|H_2)$ respectively. The prior
probabilities for the hypotheses are $p(H_1)=1-p$ and $p(H_2)=p$
where $p$ is defined in Section~\ref{sec: sysmodel}.

Now, for developing our algorithm, we assume the following three
main assumptions:

\textbf{Assumption 1}: ($y_i-\hat{y}_i$) and ($y_j-\hat{y}_j$) are
assumed to be uncorrelated for $i\neq j$.

\textbf{Assumption 2}: The noise term $v_j$ is uncorrelated of the
error $(y_i-\hat{y}_i)$ for $i\neq j$.

\textbf{Assumption 3}: The term $\gamma_j$ in (\ref{eq: gamma})
has a Gaussian distribution.

Strictly speaking, Assumption 1 is not mathematically true,
because the estimated value of one coefficient clearly influences
the estimation of the other coefficients. However, in the
following, this assumption provides a first order approximation of
a sequence of thresholds for the algorithm, instead of using a
heuristically predetermined sequence of thresholds as done in
IDE~\cite{AminBJ06}. More precisely, in the following, this
assumption is used only for deriving (\ref{eq: sigmapar}). On the
other hand, heuristically, we expect that, as the algorithm
converges to the true solution, the outputs are closer to the true
estimated values and the estimation of each coefficient has less
influence to estimating the other ones. We will study this
heuristic in our simulations (see Fig.~\ref{figerr}). In fact,
(\ref{eq: sigmapar}) requires Assumption~2, too which is just used
here. Consequently, the experiment of Fig.~\ref{figerr} will
experimentally study both Assumptions~1 and~2.



Moreover, Assumption~3 is not strictly true, too. However, since
$\gamma_j=\sum_{i=1, i\neq j}^m(y_i-\hat{y}_i)b_{ij}+v_j$ is a sum
of many (especially for large $m$'s) random variables, one expects
from Central Limit Theorem (CLT) that this assumption be a good
approximation. We will also study the validity of this assumption
experimentally (see Fig.~\ref{figkur}). This assumption will be
used in deriving the activity measure.




Now, let $\sigma^2_{\gamma_j}$ denote the variance of $\gamma_j$
which is assumed to be a Gaussian random variable by Assumption 3.
Therefore, the activity condition writes:
\begin{equation}
\label{eq: hyp3}
\begin{split}
\frac{(1-p)}{\sqrt{2\pi(\sigma^2_{\gamma_j}+\sigma^2_r)}}\exp(\frac{-(z_j-c_j)^2}{2(\sigma^2_{\gamma_j}+\sigma^2_r)})>\\
\frac{p}{\sqrt{2\pi\sigma^2_{\gamma_j}}}\exp(\frac{-(z_j-c_j)^2}{2\sigma^2_{\gamma_j}}).
\end{split}
\end{equation}
Simplifying (\ref{eq: hyp3}) with the assumption that the
(unknown) parameters $p$, $\sigma_r$ and $\sigma_{\gamma_j}$ are
known, leads to the following decision rule for the hypothesis
testing:
\begin{equation}
\label{eq: HBP}
\text{Activity}(y_j)\triangleq|z_j-c_j|>\mathrm{Th}_j.
\end{equation}
where $\mathrm{Th}_j$ is the threshold defined as:
\begin{equation}
\label{eq: th}
\mathrm{Th}_j\triangleq\frac{\sigma_{\gamma_j}}{\sigma_r}\sqrt{2(\sigma^2_r+\sigma^2_{\gamma_j})\ln(\frac{p}{1-p}\frac{\sqrt{\sigma^2_r+\sigma^2_{\gamma_j}}}{\sigma_{\gamma_j}})
}.
\end{equation}

The decision rule and the activity function in (\ref{eq: HBP}) are
the same as in IDE algorithm \cite{AminBJ06}, where one uses a
predefined decreasing sequence of thresholds. Improvements with
respect to IDE method is that the value of threshold is obtained
mathematically with respect to the parameters of the statistical
model, \ie following a Bayesian hypothesis test. Another important
difference is that, IDE only uses the same threshold for all
coefficients, while BHTA could use a different threshold for each
coefficient. However, as we will state in Section~\ref{sec:
simresult}, we use the same threshold for all the coefficients to
simplify the algorithm.

Although (\ref{eq: th}) determines the optimal threshold, it
depends on unknown parameters ($p$, $\sigma_r$ and
$\sigma_{\gamma}$) which should be estimated from the original
signal ($\xb$). Since estimating the parameters needs also the
activity vector $\qb$ which is derived itself by the value of
threshold, we use an iterative algorithm. To estimate the
parameters $p$, $\sigma_r$ and $\sigma_e$, we can use sample
estimate formulas, which are:
\begin{equation}
\label{eq: estp} \hat{p}=\frac{||\qb||_0}{m},
\end{equation}
\begin{equation}
\label{eq: estsige}
\hat{\sigma}_e=\frac{||\xb-\Phib\hat{\yb}||_2}{\sqrt{n}},
\end{equation}
\begin{equation}
\label{eq: estsigr} \hat{\sigma_r}=\frac{||\rb||_2}{\sqrt{m}}.
\end{equation}
where $\qb$ is obtained from the previous iteration of the
decision rule (\ref{eq: HBP}). In \cite{ZayyBJ08}, it has been
proved that these estimates are the MAP estimation of these
parameters knowing all other parameters. The initialization of
these parameters is also detailed in Section~\ref{sec: sim1}.

The problem here is to estimate the parameter $\sigma_{\gamma_j}$
which is the standard deviation of $\gamma_j$ in (\ref{eq:
gamma}). By taking the variance from (\ref{eq: gamma}), since
$v_j$ is a Gaussian random variable with the same variance as
$e_j$ which is equal to $\sigma^2_e$, then by Assumption 1 and
Assumption 2, we will have:
\begin{equation}
\label{eq: sigmapar} \sigma^2_{\gamma_j}=\sigma^2_e+\sum_{i=1,
i\neq j}^mb^2_{ij}\sigma^2_{i,e_y}.
\end{equation}
where $\sigma^2_{i,e_y}$ is the variance of the error term
$(y_i-\hat{y}_i)$. The accuracy of the above formula depends on
the validity of Assumptions 1 and 2, and will be experimentally
studied in Section~\ref{sec: simresult}.

If the algorithm converges, we expect that $\sigma^2_{\gamma_j}$
decreases. So, we enforce the error variance to decrease
geometrically:
\begin{equation}
\label{eq: alphapar}
\sigma^{(n+1)}_{i,e_y}=\alpha\sigma^{(n)}_{i,e_y}.
\end{equation}
where the parameter $\alpha$, less than but close to 1, determines
the rate of convergence.

In Appendix~\ref{app1}, it is shown that if we choose the minimum
\ltwo-norm solution for the first iteration, then the initial
estimate of the variance $\sigma^{2^{(0)}}_{j,e_y}$ is:
\begin{equation}
\label{eq: initpar}
\sigma^{2^{(0)}}_{j,e_y}=\sigma^2_r(\sum_{i\in\mathrm{supp}(\yb)}\psi^2_{ji})+\sigma^2_e\sum_{i}l^2_{ji}.
\end{equation}
where $\mathbf{\Psi}=[\psi_{ij}]\triangleq-\Ib+\Phib^\dagger\Phib$
and $\Lb=[l_{ij}]\triangleq 2\Phib^T-\Phib^\dagger$. The notation
$\mathrm{supp}(\yb)$ denotes the indices where the coefficients
are nonzero. But, we do not know in advance where the nonzero
elements are. So, we replace the first right side term of
(\ref{eq: initpar}) by its mathematical expectation. The
expectation $E(\sum_{i\in\mathrm{supp}(\yb)}\psi^2_{ji})$ is equal
to $E(\sum_{i}q_i\psi^2_{ji})=\sum_{i}E(q_i)\psi^2_{ji}$ where
$q_i$ is the activity of the $i$'th element which is a Bernoulli
variable, and hence $E(q_i)=(1-p)$. So, the following formula can
be used to estimate the initial parameter estimation:
\begin{equation}
\label{eq: initpar1}
\sigma^{2^{(0)}}_{j,e_y}\approx\sigma^2_r(1-p)||\boldsymbol{\psi}_j||^2_2+\sigma^2_e\sum_{i}l^2_{ji}.
\end{equation}
where $\boldsymbol{\psi}_j$ is the $j$'th row of the matrix
$\boldsymbol{\Psi}$.

From (\ref{eq: initpar1}), (\ref{eq: sigmapar}) and the assumption
that the error variances $\sigma^2_{i,e_y}$ tends to zero at final
iterations, we can find that the value of $\sigma^2_{\gamma_j}$
varies from a large initial value
$\sigma^{2^{(0)}}_{\gamma_j}=\sigma^2_e+\sum_{i=1, i\neq
j}^mb^2_{ij}\sigma^{2^{(0)}}_{i,e_y}$ to a small value
$\sigma^{2^{(\infty)}}_{\gamma_j}=\sigma^2_e$. So, from (\ref{eq:
th}), the threshold is changed from an initial large value
$\mathrm{Th}_j^{(0)}\triangleq\mathrm{Th}|_{\sigma^{(0)}_{\gamma_j}}$
to a small final value
$\mathrm{Th}_j^{(\infty)}\triangleq\mathrm{Th}|_{\sigma^{{(\infty)}}_{\gamma_j}}$.
The initial value and the final value (after infinite iterations)
of the threshold are:
\begin{equation}
\label{eq: thi}
\mathrm{Th}_j^{(0)}=\mathrm{Th}|_{\sigma^{(0)}_{\gamma_j}},
\end{equation}
\begin{equation}
\label{eq: thf}
\mathrm{Th}_j^{(\infty)}=\mathrm{Th}^{(\infty)}=\mathrm{Th}|_{\sigma_{\gamma_j}=\sigma_e}\approx
K\sigma_e.
\end{equation}
where $K=\sqrt{2\ln(\frac{p}{1-p}\frac{\sigma_r}{\sigma_e})}$. In
(\ref{eq: thf}), it has been assumed that the algorithm converges
to the true solution and hence
$\sigma_{\gamma_j}\rightarrow\sigma_e$. As we can see from
(\ref{eq: thi}), the initial thresholds are different for each
coefficient. But, all the thresholds are converging to the same
value (\ref{eq: thf}).

As we explain in Section~\ref{sec: simresult}, a common threshold
is used for all coefficients for simplicity of the algorithm. As
the value of threshold changes from a large value to a small
value, the algorithm can detect more and more atoms. During the
first iterations, the optimal thresholding strategy in (\ref{eq:
th}) changes the thresholds very fast and then after a few
iterations, the thresholds converge to the final small value.


In the thresholding strategy (\ref{eq: th}), although there can be
a simple stopping rule for iterations based on the value of
thresholds (which will be explained in our experiments), the
number of required iterations for convergence are not known in
advance. So, we can use another threshold to predict the number of
the iterations in advance. The simplest way for updating the
threshold is to decrease the threshold geometrically from the
initial value $\mathrm{Th}_j^{(0)}$ in (\ref{eq: thi}) to final
value $\mathrm{Th}_j^{(\infty)}$ in (\ref{eq: thf}):
\begin{equation}
\label{eq: thsimple}
\mathrm{Th}^{*(n+1)}_j=\alpha\mathrm{Th}^{*(n)}_j.
\end{equation}
where superscript $*$ is for simple thresholding strategy. Since
the final threshold is the same for all coefficients, we should
also use the same threshold for initialization in the simple
thresholding strategy. As we see in Section~\ref{sec: simresult},
we also use the same thresholds for optimal thresholding. So, we
can use the same initial value for threshold in simple
thresholding just like in optimal thresholding (\ie
$\mathrm{Th}_j^{(0)}=\mathrm{Th}^{(0)}$). So, in simple
thresholding, the required number of iterations is:
\begin{equation}
\label{eq: tnum}
t=\frac{\ln(\frac{\mathrm{Th}^{(\infty)}}{\mathrm{Th}^{(0)}})}{\ln(\alpha)}.
\end{equation}
where $\mathrm{Th}^{(\infty)}$ is as defined in (\ref{eq: thf}).
In other words, using $t$ iterations of (\ref{eq: thsimple}),
$\mathrm{Th}^{*(n)}_j$ changes from $\mathrm{Th}^{(0)}$ to
$\mathrm{Th}^{(\infty)}$ which were defined in (\ref{eq: thi}) and
(\ref{eq: thf}). So, with this strategy of selecting the
thresholds, we can predict the number of iterations in advance.
The practical choice will be explained in the experimental results
section. We will refer to this method as simple thresholding,
while the straightforward method is referred to as optimal
thresholding. The simple thresholding strategy is similar to IDE
with the difference that here we know the first and last values of
thresholds while IDE has no ideas for initial and last values of
thresholds.


After updating the activity vector based on decision rule in
(\ref{eq: HBP}), the estimation of amplitude vector $\rb$ which
was defined in Section~\ref{sec: sysmodel}, based on this
estimated activity vector can be done by a Linear Least Square
(LLS) estimation \cite{Bjor96},\cite{ZayyBJ07DSP}:
\begin{equation}
\label{eq: Estep}
\bf{\hat{\rb}}=\sigma_r^2\bf{\hat{\Qb}}\Phib^T(\sigma_r^2\Phib\hat{\Qb}\Phib^T+\sigma_e^2\Ib)^{-1}\xb.
\end{equation}
where $\hat{\qb}$ is the estimated activity vector and
$\hat{\Qb}=\diag(\hat{\qb})$. It is worth mentioning that
(\ref{eq: Estep}) has the same type of update as used in iterative
re-weighted least squares algorithms such as FOCUSS algorithm
\cite{GoroR97}. In fact, (\ref{eq: Estep}) is nothing but this
standard approach, but with a novel way for calculating the
weights.

\subsection{Soft-BHTA}
\label{sec: SBPA} In the previous subsection, we presented the
Hard-version of BHTA. As we saw, determining the threshold is
relatively complex. Therefore, in this section we suggest a
Soft-version of BHTA to avoid the threshold computation. The main
idea of soft version of BHTA is to use soft posterior
probabilities as the soft hypothesis testing results instead of
binary values `0' or `1' for activity measure $q_i$. If the value
of the posterior probability $p(q_i=1|z_i)$ is high, then it means
that it is more probable that the $i$'th coefficients are nonzero
or the $i$'th atom is active. So, we simply replace $q_i$ by
$p(q_i=1|z_i)$ as we will see at (\ref{eq: softth}). At the final
iteration, we use a hard thresholding for providing a binary value
for each $q_i$. So, with this trick, all atoms are participated in
the sparse representation in the initial iterations. Then, we
replace the activity measures by the posteriors, which determine
active (or inactive) atoms of the sparse representation. In this
case, the $z_i$'s are the correlations which are used in pursuit
algorithms and $p(q_i=1|z_i)$ is the posterior of the $i$'th
coefficient conditionally to the correlations.

To compute the posteriors, we use the Bayes rule as:
\begin{equation}
p(q_i=1|z_i)=p(H_1|z_i)=\frac{p(H_1)p(z_i|H_1)}{p(H_1)p(z_i|H_1)+p(H_2)p(z_i|H_2)}.
\end{equation}
Using (\ref{eq: hyp}), in each iteration of soft version of BHTA,
we must do the following update for the soft-activity measure:
\begin{equation}
\label{eq: softth}
q_i\gets\frac{\frac{(1-p)}{\sqrt{\sigma^2_r+\sigma^2_{\gamma}}}\exp(\frac{-(z_i-m_i)^2}{2(\sigma^2_r+\sigma^2_{\gamma})})}{\frac{(1-p)}{\sqrt{\sigma^2_r+\sigma^2_{\gamma}}}\exp(\frac{-(z_i-m_i)^2}{2(\sigma^2_r+\sigma^2_{\gamma})})+\frac{p}{\sigma_{\gamma}}\exp(\frac{-(z_i-m_i)^2}{2\sigma^2_{\gamma}})}.
\end{equation}
where parameters $p$, $\sigma_r$ and $\sigma_{\gamma}$ are
obtained and updated as in the hard version of BHTA. After the
convergence, we can use a simple hard thresholding as
$p(q_i=1|z_i)\ge 0.5$ to obtain the active atoms.

After updating the activity, we can use a formula similar to
(\ref{eq: Estep}) for updating the amplitude vector.

\subsection{Summary}
Finally, to summarize the presentation of BHTA, we represented the
detailed Hard-BHTA algorithm in Fig.~\ref{fig6}. Initialization is
done by minimum \ltwo-norm solution (see (\ref{eq: ltwo}) in
Appendix~\ref{app1}). Updating the activity vector in Hard-BHTA is
done by the decision rule in (\ref{eq: HBP}). Updating the
coefficients or amplitudes is done by (\ref{eq: Estep}).
Similarly, updating the parameters are done by equations (\ref{eq:
sigmapar}), (\ref{eq: alphapar}), (\ref{eq: initpar1}) and with
parameters $p$, $\sigma_e$ and $\sigma_r$ computed according to
(\ref{eq: estp}), (\ref{eq: estsige}) and (\ref{eq: estsigr}).
Threshold determination in Hard-BHTA is done by equations
(\ref{eq: initpar1}), (\ref{eq: alphapar}) and then (\ref{eq:
sigmapar}) for the variance of coefficient errors. Then, as we
explained in Section~\ref{sec: HBPA}, we suggested two different
thresholding strategy which are optimal thresholding (\ref{eq:
th}) and simple thresholding (\ref{eq: thsimple}). We investigate
these different strategies in the simulation results.

\begin{figure}
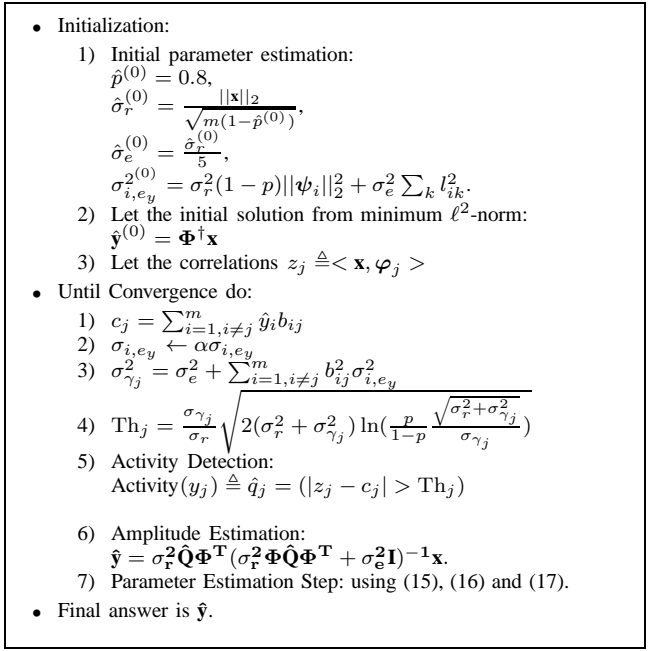

  \centering \vrule
    \begin{minipage}{8.5cm} 
    \hrule \vspace{0.5em} 
    \begin{minipage}{7.5cm} 

      {
        \footnotesize
        \def\baselinestretch{1}

        \begin{itemize}
        \item Initialization:

           \begin{enumerate}
             \item Initial parameter estimation:

             $\hat{p}^{(0)}=0.8$,\\
             $\hat{\sigma}_r^{(0)}=\frac{||\xb||_2}{\sqrt{m(1-\hat{p}^{(0)})}}$,\\
             $\hat{\sigma}_e^{(0)}=\frac{\hat{\sigma}_r^{(0)}}{5}$,\\
             $\sigma^{2^{(0)}}_{i,e_y}=\sigma^2_r(1-p)||\boldsymbol{\psi}_i||^2_2+\sigma^2_e\sum_{k}l^2_{ik}$.

             \item Let the initial solution from minimum \ltwo-norm:

             $\hat{\yb}^{(0)}=\Phib^\dagger\xb$
             \item Let the correlations $z_j\triangleq<\xb,\boldsymbol{\phi}_j>$

           \end{enumerate}

        \item Until Convergence do:
          \begin{enumerate}
             \item $c_j=\sum_{i=1, i\neq j}^m\hat{y}_ib_{ij}$
             \item $\sigma_{i,e_y}\leftarrow\alpha\sigma_{i,e_y}$
             \item $\sigma^2_{\gamma_j}=\sigma^2_e+\sum_{i=1, i\neq j}^mb^2_{ij}\sigma^2_{i,e_y}$
             \item $\mathrm{Th}_j=\frac{\sigma_{\gamma_j}}{\sigma_r}\sqrt{2(\sigma^2_r+\sigma^2_{\gamma_j})\ln(\frac{p}{1-p}\frac{\sqrt{\sigma^2_r+\sigma^2_{\gamma_j}}}{\sigma_{\gamma_j}})}$
             \item Activity Detection: \\
             \quad $\text{Activity}(y_j)\triangleq \hat{q}_j=(|z_j-c_j|>\mathrm{Th}_j)$\\
             \item Amplitude Estimation:\\
             \quad $\bf{\hat{\yb}}=\sigma_r^2\bf{\hat{\Qb}}\Phib^T(\sigma_r^2\Phib\hat{\Qb}\Phib^T+\sigma_e^2\Ib)^{-1}\xb.$

             \item Parameter Estimation Step: using (\ref{eq:
             estp}), (\ref{eq: estsige}) and (\ref{eq: estsigr}).

          \end{enumerate}
        \item Final answer is $\bf{\hat{\yb}}$.
        \end{itemize}
      }
    \end{minipage}
    \vspace{1em} \hrule
  \end{minipage}\vrule \\
\caption{The Hard-BHTA algorithm.} \label{fig6}
\end{figure}

%
%
%
%
%
%
%
%
%
%
%
%
%
%
%

\section{Stability Analysis}
\label{sec: conv}


Because of the thresholding strategy, a complete convergence
analysis to the algorithm is very tricky, and is not addressed in
this paper. Hence, in this section, we only study the stability of
BHTA, \ie if the algorithm does not diverge and is stable.


The stability of the BHTA is equivalent to the convergence of the
sequence of thresholds in (\ref{eq: th}). As we know, this is a
positive sequence and it is bounded below by zero. So, if this
sequence is a decreasing sequence, then it converges of course to
the value in (\ref{eq: thf}). In Appendix~\ref{app2}, we show that
with the assumption that $\sigma_{\gamma_j}\ll \sigma_r$, the
sufficient condition for the convergence of the sequence of $j$'th
threshold is:
\begin{equation}
\label{eq: suff} \frac{\sigma^2_r}{\sigma^2_e}>\frac{\sum_{i=1,
i\neq j}^mb^2_{ij}\sum_rl^2_{jr} }{(\frac{p}{(1-p)e})^2-\sum_{i=1,
i\neq j}^mb^2_{ij} \sum_{r\in\mathrm{supp}(\yb)}\psi^2_{jr}}.
\end{equation}
where $e$ is the neper number,
$\Bb=\boldsymbol{\Phib}^T\boldsymbol{\Phib}$,
$\boldsymbol{\Psi}=-\Ib+\boldsymbol{\Phib}^\dagger\boldsymbol{\Phib}$
and $\Lb=[l_{ij}]\triangleq 2\Phib^T-\Phib^\dagger$. If we define
the input SNR as in (\ref{eq: snrdef1}), then (\ref{eq: suff}) is
equivalent to have an input SNR greater than a minimum input SNR,
\ie $\mbox{SNR}_{i}>\mbox{SNR}_{min}$ which is:
\begin{equation}
\mbox{SNR}_{min}(\yb,j)\triangleq 10\log\frac{\sum_{i=1, i\neq
j}^mb^2_{ij}\sum_rl^2_{jr} }{(\frac{p}{(1-p)e})^2-\sum_{i=1, i\neq
j}^mb^2_{ij} \sum_{r\in\mathrm{supp}(\yb)}\psi^2_{jr}}.
\end{equation}
where this minimum SNR depends on the unknown $\yb$. For canceling
the dependence on $\yb$, we replace the denominator by its
expectation with respect to $\yb$ (like for deriving (\ref{eq:
initpar1})) and the minimum SNR becomes:
\begin{equation}
\label{eq: minsnr} \mbox{SNR}_{min}(j)\triangleq
10\log\frac{||\boldsymbol{\ell}^T_j||^2_2(||\bb_j||^2_2-1)}{K+(1-p)||\boldsymbol{\psi}_j||^2_2(||\bb_j||^2_2-1)}.
\end{equation}
where $K=(\frac{p}{(1-p)e})^2$, $\bb_j$ is the $j$'th column of
$\Bb$ and $\boldsymbol{\ell}^T_j$ is the $j$'th row of $\Lb$.
Although the formula for minimum input SNR in (\ref{eq: minsnr})
seems complicated, it is not very restrictive, \ie this minimum
value is not very high. To evaluate the values for the minimum
input SNR, we compute them for the practical case of CS where the
real signals are sparse in DCT domain. In this case, the matrix
$\Phib=\boldsymbol{\Gamma}\Db$ where $\boldsymbol{\Gamma}$ is the
random CS measurement matrix and $\Db$ is the DCT matrix. The
random measurement matrix elements are drawn from a zero mean
normal random distribution with unit variance. The columns of the
dictionary matrix $\Phib$ are normalized to have unit norms. A
typical simulation in this case shows that the minimum and maximum
values of $\mbox{SNR}_{min}(j)$ over different $j$'s are -11.4127
dB and 0.3560 dB. So, the practical values of $\mbox{SNR}_{min}$
are not very high, and the sufficient condition for stability
(\ref{eq: suff}) is a weak condition and easily satisfied.


For the stability analysis of soft-BHTA, we define the two terms
in (\ref{eq: softth}) as
$l_1\triangleq\frac{(1-p)}{\sqrt{\sigma^2_r+\sigma^2_{\gamma}}}\exp(\frac{-(z_i-m_i)^2}{2(\sigma^2_r+\sigma^2_{\gamma})})$
and
$l_2\triangleq\frac{p}{\sigma_{\gamma}}\exp(\frac{-(z_i-m_i)^2}{2\sigma^2_{\gamma}})$.
With these definitions, $l_1\ge l_2$ is equivalent to $q_i\ge 0.5$
and $l_1<l_2$ is equivalent to $q_i<0.5$. It is simple to show
that the condition $l_1\ge l_2$ is equivalent to $|z_i-m_i|\ge
\mathrm{Th}_i$ which is similar to the decision rule of Hard-BHTA
in (\ref{eq: HBP}). Therefore, the stability conditions for
hard-BHTA and soft-BHTA are the same.

\section{Experiments}
\label{sec: simresult} The BHTA algorithm is investigated in this
section with three different categories of simulation. First, in
subsection~\ref{sec: sim1}, we only consider soft and hard
versions of the BHTA algorithm. It includes the two different
thresholding strategies and some detailed implementation issues of
the algorithm. Secondly, in Section~\ref{sec: sim2}, comparison
will be done with the other main algorithms for sparse
representation both from complexity and estimation accuracy
viewpoints. The performance of the algorithms is compared using
the Signal to Noise Ratio between the true coefficients and the
recovered coefficients, which is defined as:
\begin{equation}
\label{eq: snrdef}
\mbox{SNR}_o\triangleq10\log(\frac{||\yb||_2}{||\yb-\hat{\yb}||_2}).
\end{equation}
where the index $o$ denotes output SNR. In fact, this SNR in the
coefficient domain determines the capability of the sparse
representation algorithm to recover the true sparse coefficients
in average. We define another measure which determines the noise
level. We refer to it as input SNR:
\begin{equation}
\label{eq: snrdef1}
\mbox{SNR}_i\triangleq20\log(\frac{\sigma_r}{\sigma_e}).
\end{equation}
This input SNR is varied from 20dB to 50dB in the experiments.

We use the CPU time as a measure of complexity. Although, the CPU
time is not an exact measure, it can give us a rough estimation of
the complexity for comparing our algorithms. Our simulations were
performed in MATLAB7.0 environment using an AMD Athlon Dual core
4600 with 896 MB of RAM and under Windows XP operating system.

\subsection{Implementation issues of BHTA}
\label{sec: sim1} In this part of our experiments, the
implementation aspects of the BHTA algorithm is experimentally
discussed and evaluated. We mainly have two hard and soft versions
of the BHTA algorithm, since we use two distinct methods for
updating the threshold.

We used a random dictionary matrix with normalized columns whose
entries are previously drawn according to a uniform distribution
in $[-1,1]$. The number of atoms is set to $m=512$ and the signal
length to $n=256$. For the sparse coefficients, we used the model
(\ref{eq: spiky}) with the probability $p=0.9$ and unit variance
for the active coefficients ($\sigma_r=1$). So, on the average,
about 51 atoms are active in the sparse representation of the
signal. The noises or errors are Gaussian with zero-mean and
different variances. The measure of performance, the output SNR
(\ref{eq: snrdef}), is averaged over 100 different random
realizations of the dictionary, sparse coefficients and noise
vector.

For simplifying the algorithm, we use the same variance and
threshold for all coefficients. This simplification reduces some
of our calculations by a factor of $\frac{1}{m}$. Since the value
of $\sigma^2_e$ is not known in advance and the term
$\sigma^2_e\sum_{i}l^2_{ji}$ is small in comparison to other term,
we select
$\sigma^{2^{(0)}}_{j,e_y}\approx\sigma^2_r(1-p)||\boldsymbol{\psi}_j||^2_2$.
To remove the dependency on the index $j$, we select
$\sigma^{2^{(0)}}_{j,e_y}\approx\sigma^2_r(1-p)||\boldsymbol{\Psi}||^2_F$
where the approximation
$||\boldsymbol{\psi}_j||^2_2\approx\frac{||\boldsymbol{\Psi}||^2_F}{m}$
is assumed for large random matrix $\boldsymbol{\Psi}$. With this
initialization (which is independent of the coefficient index) and
(\ref{eq: alphapar}), all the error variances $\sigma_{i,e_y}$ are
independent of the index $i$ and assumed to be $\sigma_{e_y}$. So,
(\ref{eq: sigmapar}) will reduce to
$\sigma^2_{\gamma_j}=\sigma^2_e+\sigma^2_{e_y}(||\bb_j||^2_2-1)$.
To omit the dependency on $j$, we use
$\sigma^2_{\gamma}=\sigma^2_e+\beta\sigma^2_{e_y}$ where
$\beta\approx\frac{||\Bb||^2_F}{m}-1$ (since
$||\bb_j||^2_2\approx\frac{||\Bb||^2_F}{m}$ is a first order
approximation of $E\{||\bb_j||^2_2\}=E\{\frac{||\Bb||^2_F}{m}\}$
which holds for random dictionaries. Finally, the values of
thresholds are the same for all the coefficient indexes.

The initial values of the unknown statistical parameters ($p$,
$\sigma_r$ and $\sigma_e$) are $\hat{p}^{(0)}=0.8$,
$\hat{\sigma}^{(0)}_r=\frac{||\xb||_2}{\sqrt{m(1-\hat{p}^{(0)})}}$
and $\hat{\sigma}^{(0)}_e=\frac{\hat{\sigma}^{(0)}_r}{5}$ which is
similar to the initialization used in \cite{ZayyBJ08},
\cite{ZayyBJ08ICASSP}. We can propose some stopping rules for
Hard-BHTA. In Hard-BHTA, we used
$\mathrm{Max}(\mathrm{|\mathbf{Th}}^
{(n+1)}-\mathrm{\mathbf{Th}}^{(n)}|)<\frac{\hat{\sigma}_r}{1000})$
as an stopping rule. For Soft-BHTA, a similar stopping rule is
$\mathrm{Max}(|\qb^{(n+1)}-\qb^{(n)}|)<\frac{1}{100}$.

\begin{figure}[tb]
\begin{center}
\includegraphics[width=8cm]{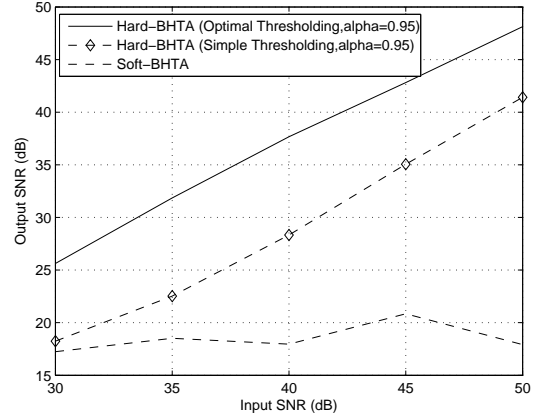}
\end{center}
\caption{\footnotesize The output SNR averaged on 100 runs versus
the input SNR for Hard-BHTA with two different thresholding
strategies and Soft-BHTA. The parameters are $m=512$, $n=256$,
$p=0.9$, $\sigma_r=1$, $\alpha=0.95$.} \label{fig7}
\end{figure}

For the simple thresholding strategy (which is very similar to
IDE), we start from the initial threshold $\mathrm{Th}^{(0)}$ to
the final value $\mathrm{Th}^{(\infty)}$ by the geometric series
(\ref{eq: thsimple}). To compute $\mathrm{Th}^{(\infty)}\approx
K\sigma_e$, we need the value $\frac{\sigma_r}{\sigma_e}$. In the
simulations of this section, we select
$\frac{\sigma_r}{\sigma_e}=100$ for any noise levels and the
parameter $\alpha=0.95$ for both simple thresholding and optimal
thresholding. Figure~\ref{fig7} shows the results of the two
versions of Hard-BHTA (with the two thresholding strategies) and
Soft-BHTA. Clearly, performance of Hard-BHTA is much better than
Soft-BHTA performance. Of course, the optimal thresholding
strategy (\ref{eq: th}) yields better results than the simpler
strategy (\ref{eq: thsimple}).

\begin{figure}[tb]
\begin{center}
\includegraphics[width=8cm]{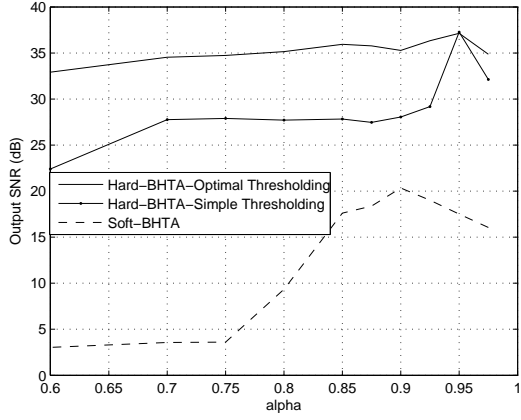}
\end{center}
\caption{\footnotesize The output SNR averaged on 100 runs versus
the simulation parameter $\alpha$. Other parameters are $m=512$,
$n=256$, $p=0.9$, $\sigma_r=1$.} \label{fig10}
\end{figure}

Finally, to determine the best value of the parameter $\alpha$ in
(\ref{eq: alphapar}) and (\ref{eq: thsimple}), we represent the
results of our algorithm with respect to the value of $\alpha$
when $\sigma_e=0.01$ in Fig.~\ref{fig10}. As it can be seen,
better results are obtained when the value is around $\alpha=0.95$
for optimal thresholding and for simple thresholding. However, we
use $\alpha=0.95$ for the next experiments unless we state
otherwise. As we can see, the Soft-BHTA and Hard-BHTA with simple
thresholding are sensitive to the value of parameter $\alpha$,
while Hard-BHTA with optimal thresholding is less sensitive to
this parameter.

\subsection{Investigating the assumptions}
\label{sec: sim3} 
\begin{figure}[tb]
\begin{center}
\includegraphics[width=8cm]{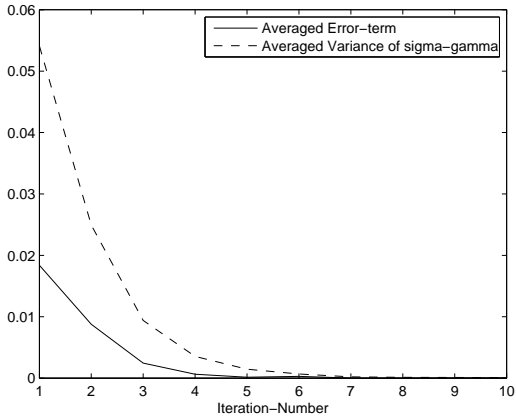}
\end{center}
\caption{\footnotesize The error term
$|\sigma^2_{\gamma_j}-\sigma^2_e-\sum_{i=1, i\neq
j}^mb^2_{ij}\sigma^2_{i,e_y}|$ and variance $\sigma^2_{\gamma_j}$
averaged over all indexes $1\le j\le m$ versus the iteration
number. It is computed over 100 runs of simulations. The
parameters are $m=512$, $n=256$, $p=0.9$, $\sigma_r=1$ and
$\sigma_n=0.01$.} \label{figerr}
\end{figure}

In these experiments, the Assumptions 1 to 3 of
Section~\ref{sec: HBPA} are investigated. Since Assumptions 1 and
2 have only been used in deriving (\ref{eq: sigmapar}), the
influence of these assumptions is experimentally investigated by
computing the absolute difference between both sides of (\ref{eq:
sigmapar}) \ie the error term
$|\sigma^2_{\gamma_j}-\sigma^2_e-\sum_{i=1, i\neq
j}^mb^2_{ij}\sigma^2_{i,e_y}|$. Figure~\ref{figerr} shows the
error term over all indices $1\le j\le m$ versus the iteration
number. It can be seen that the averaged error term is small in
comparison to $\sigma^2_{\gamma_j}$, and vanishes after a few (5
to 6) iterations. In other words, as the algorithms converges to
the solution, the Assumptions 1 and 2 become very accurate.

\begin{figure}[tb]
\begin{center}
\includegraphics[width=8cm]{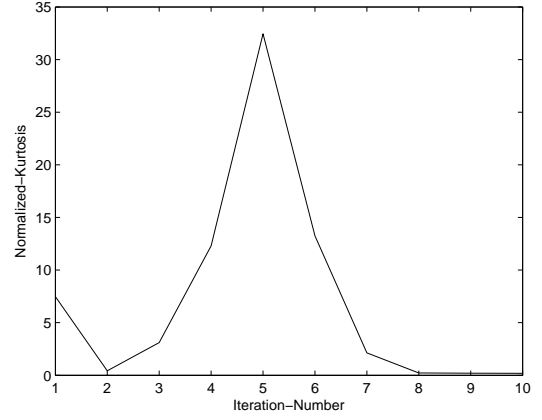}
\end{center}
\caption{\footnotesize The normalized kurtosis computed over all
indexes $1\le j\le m$ and 100 runs of simulations versus the
iteration number. The parameters are $m=512$, $n=256$, $p=0.9$,
$\sigma_r=1$ and $\sigma_n=0.01$.} \label{figkur}
\end{figure}

Assumption 3 (the Gaussianity of $\gamma_j$), is evaluated by
computing the normalized Kurtosis defined as
$\mathrm{Kurt}(\gamma_j)\triangleq\frac{\gamma^4_j}{E^2(\gamma^2_j)}-3$
\cite{HyvaKO01}. Recall that the kurtosis of a Gaussian random
variable is zero. We averaged this measure over all coefficients
indexes $j$ and also over runs of simulation. Averaged kurtosis
versus iteration number is showed in Fig.~\ref{figkur}. It can be
seen that the value of kurtosis is small after some iterations and
hence the assumption of Gaussianity of $\gamma_j$ would be a good
approximation after a few iterations.

\subsection{Comparison with other sparse representation algorithms}
\label{sec: sim2} In this experiment, we only compare the optimal
thresholding version of Hard-BHTA and Soft-BHTA with other main
sparse representation algorithms such as BP, MP, OMP, StOMP, SL0,
BCS, GP, GPSR and IBA. In this experiment, we use another model
for generating the sparse coefficients. We choose the inactivity
probability $p=0.9$ and all active coefficients are set equal to 1
instead to be distributed as a Gaussian random variable with a
unit variance. The locations of active coefficients are uniformly
random. The input SNR is defined as
$\mbox{SNR}_i\triangleq20\log(\frac{1}{\sigma_e})$. The
comparisons are done in three cases. The first case is the
comparison of the average estimation accuracy (Output SNR) versus
the input noise level (Input SNR). The second comparison is the
same measure of estimation accuracy (Output SNR) versus the
sparsity level. Finally, we compare complexity of the different
algorithms. In all experiments, the results are averaged over 100
different runs, with random dictionary and random sparse
coefficents.

\begin{figure}[tb]
\begin{center}
\includegraphics[width=9cm]{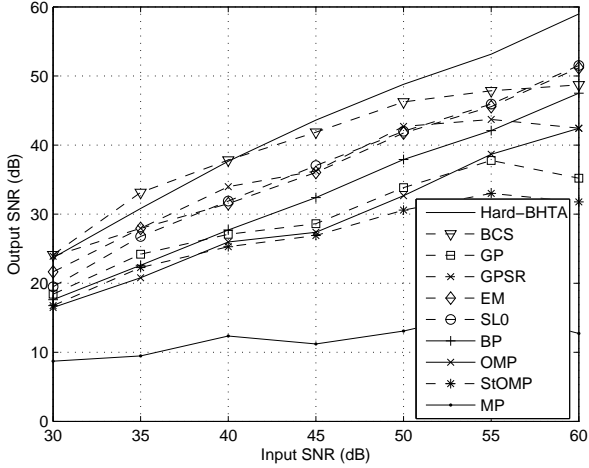}
\end{center}
\caption{\footnotesize The averaged output SNR on 100 runs of
simulations versus input SNR for various algorithms. The
parameters are $m=512$, $n=256$, $p=0.9$, $\alpha=0.95$.}
\label{fig11}
\end{figure}

For BHTA, we use the simulation parameters used in the previous
experiment. BP algorithm was tested using \lone-magic package
\cite{CandR05}. Since there are 51 active atoms in average, we run
the MP, OMP and StOMP algorithms (implemented by
SparseLab\footnote{The codes SolveMP, SolveOMP, SolveStOMP.m are
available at http://sparselab.stanford.edu}) for twice the number
of active atoms which is 102 (a similar strategy is used in
\cite{BlumD081} for yielding better performances). For StOMP, we
used default parameters of the SparseLab code with the difference
that we used similar number of iterations to MP and OMP (instead
of 10 which is the default value). For SL0 algorithm\footnote{The
code sl0.m is used which is available at
http://ee.sharif.edu/\~{}SLzero}, we used the minimum $\sigma$
equal to $\sigma_e$ and the decreasing factor, a parameter which
determined a tradeoff between accuracy and speed, equal to 0.9.
For the IBA algorithm
, we used 4 iterations for both the M-step and the overall
algorithm \cite{ZayyBJ08}. For GPSR algorithm\footnote{The code
GPSR\_fun.m is used which is available at
http://www.lx.it.pt/\~{}mtf/GPSR/GPSR6.0} \cite{FiguNW07}, we used
$\tau=0.1||\boldsymbol{\Phi}^T\xb||_{\infty}$ as suggested by the
authors. We also use a debiasing step in GPSR algorithm. The
algorithm stops if the norm of the difference between two
consecutive estimates, divided by the norm of one of them falls
below $10^{-4}$. The other parameters of GPSR are the default
values. We also used the recommended and default parameters for
BCS\footnote{The codes in bcs-vb.zip are used which are available
at http://people.ee.duke.edu/\~{}lihan/cs} \cite{JiXC08}. We used
Sparsify toolbox for GP algorithm\footnote{We used the latest
version of code greed\_gp.m, available at
http://www.see.ed.ac.uk/\~{}tblumens/sparsify/sparsify.html}
\cite{BlumD081}, with default parameters and we stop the algorithm
if the mean squared error of residual is below $\sigma^2_e$.
Figure~\ref{fig11} shows the performance of the various algorithms
(output SNR in coefficient domain) versus the noise level (input
SNR). It shows that our algorithm is one of the best algorithms in
terms of estimation accuracy specially for low noises.

\begin{figure}[tb]
\begin{center}
\includegraphics[width=9cm]{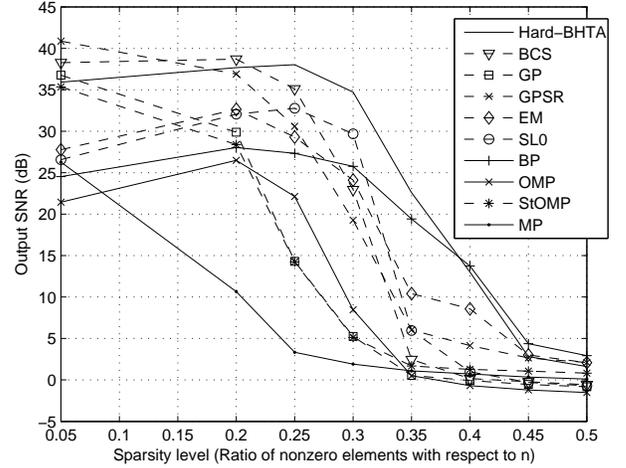}
\end{center}
\caption{\footnotesize The averaged Output SNR on 100 runs of
simulations versus sparsity level. The parameters are $m=512$,
$n=256$, $\sigma_e=0.01$, $\alpha=0.95$.} \label{fig12}
\end{figure}

To investigate the performance of the algorithms for various
sparsity levels, we plot (Fig.~\ref{fig12}) the output SNR versus
sparsity level which is determined in our statistical model
(\ref{eq: spiky}) by probability $(1-p)$. In this experiment, we
used a fixed number of nonzero coefficients with amplitudes equal
to 1. The sparsity ratio is defined as $\frac{||y||_0}{n}$. Again,
it can be seen in Fig.~\ref{fig12} that the Hard-BHTA algorithm is
one of the best algorithms.

\begin{figure}[tb]
\begin{center}
\includegraphics[width=9cm]{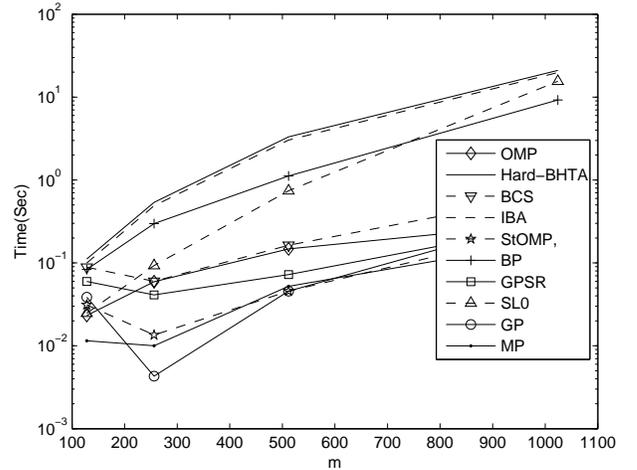}
\end{center}
\caption{\footnotesize The averaged simulation time on 100 runs of
simulations versus $m$ in the case of $m=2n$. The parameters are
$p=0.9$, $\sigma_e=0.01$, $\alpha=0.95$.} \label{fig13}
\end{figure}

Finally, we compare the algorithm in terms of speed.
Figure~\ref{fig13} shows the average simulation time of various
algorithms with respect to the dimension of our sparse
representation problem (\ie signal length). The dimension of our
problem is determined with the number of atoms and the length of
the signal. In this experiment, we used $m=2n$ for different
signal lengths from 64 to 512. It shows that our algorithm is the
most complex method.


\subsection{Comparison of algorithms in real-field decoding application}
\label{eq: decoding} In this section, we compare the algorithms in
real-field coding. In real-field coding, we first encode a block
vector of real-valued samples by a random generating matrix.
Assume the input message vector is $\sbb=[s_1,s_2,...,s_n]^T$. The
encoded message is $\xb=\Gb\sbb$ where $\Gb$ is an $n\times m$
matrix with $n<m$ (adding redundancy to input messages). Then, we
assume that channel adds both impulse errors and a background
noise. So, the channel output is equal to $\yb=\xb+\eb+\vb$ where
$\eb$ is channel errors and $\vb$ is the background noise. We can
define a parity check matrix $\Hb$ associated to the generating
matrix $\Gb$ such that $\Hb\Gb=0$ \cite{CandT05}. Then, the errors
can be reconstructed by solving the underdetermined linear system
of equations $\tilde{\yb}\triangleq\Hb\yb=\Hb\eb+\wb$ where
$\wb\triangleq\Hb\vb$ is the noise term. After estimating the
error vector $\hat{\eb}$ by means of sparse representation
algorithms, it can be subtracted from the output channel to yield
the corrected encoded message $\hat{\xb}$. Finally, the original
messages can be recovered using
$\hat{\sbb}=\Gb^{\dagger}\hat{\xb}$ where $\Gb^{\dagger}$ denotes
the pseudo-inverse of $\Gb$.

\begin{figure}[tb]
\begin{center}
\includegraphics[width=9cm]{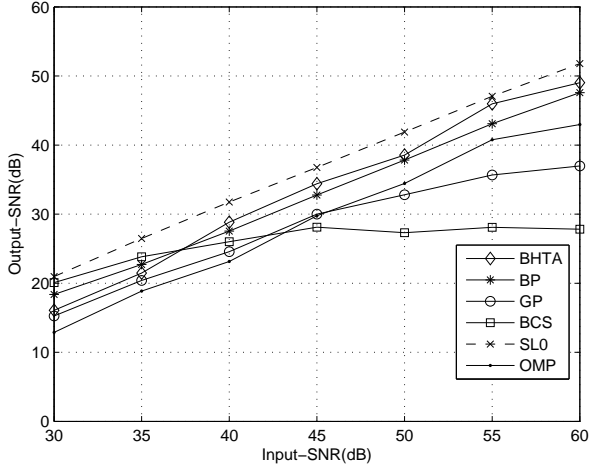}
\end{center}
\caption{\footnotesize The averaged output SNR on 100 runs of
simulations versus input SNR for various algorithms in real-field
decoding when impulse noise is BG with parameters $p=0.9$ and
$\sigma_r=1$. The other parameters are $m=256$, $n=128$,
$\alpha=0.95$.} \label{fig14}
\end{figure}

The standard Lena image is used as input message. The pixels of
image are vectorized and then divided in blocks of length $n=128$.
Entries of the generating matrix $g_{ij}$ are also randomly
selected from uniform distribution in $[-1,1]$. For channel
impulse errors, we used the model (\ref{eq: spiky}).
The background noise $\vb$ is generated from zero mean Gaussian
distribution with variance $\sigma^2_v$. The input SNR is defined
as $\mbox{SNR}_i\triangleq20\log(\frac{\sigma_r}{\sigma_v})$. The
output SNR between the original message $\sbb$ and the estimated
message $\bf{\hat{\sbb}}$ is similarly defined as
$\mbox{SNR}_o\triangleq10\log(\frac{||\sbb||_2}{||\sbb-\bf{\hat{\sbb}}||_2})$.
We vary the input SNR from 30dB to 60dB. Figure~\ref{fig14} shows
the averaged result (over 100 blocks of the Lena image) of output
SNR versus input SNR for BG model for errors. For more clarity, we
only compare our BHTA algorithm with BP, GP, BCS, SL0 and OMP, and
parameters are chosen as in the previous experiments. There are
just one difference: we used $\mbox{stopTol}=10^{-6}$ for GP
algorithm for achieving better results. The results show that
again the BHTA algorithm is one of the best algorithms for
real-field decoding application.


\section{Conclusions}
In this paper, we proposed a Bayesian hypothesis testing algorithm
for sparse representation problem which can also be used in other
contexts like CS or SCA. The main idea of this algorithm is to use
rather simple Bayesian hypothesis test to estimate which atoms are
active in the sparse expansion of the signal. The activities of
atoms, which are detected through a Bayesian test, are based on a
comparison of the activity measure with a threshold. The interest
of the Hard-BHTA algorithm is its ability to determine the
thresholds mathematically with simple parameter estimation
techniques rather than heuristically. It can be computed
practically with simple parameter estimation techniques. The
comparison of Hard-BHTA algorithm with the state of the art
algorithms shows that Hard-BHTA algorithm achieves one of the best
performances, but at the price of the highest complexity.

\appendices

\section{Initial parameter estimation for minimum \ltwo-norm solution}
\label{app1} If the minimum \ltwo-norm solution is selected as the
solution for the first iteration, then we have:
\begin{equation}
\label{eq: ltwo}
\hat{\yb}^{(0)}=\Phib^\dagger\xb=\Hb\yb+\boldsymbol{\eta}.
\end{equation}
where $\Hb\triangleq\Phib^\dagger\Phib$ and
$\boldsymbol{\eta}\triangleq\Phib^\dagger\eb$. Then, each element
of the initial solution can be written as:
\begin{equation}
\label{eq: app1} \hat{y}^{(0)}_i=\sum_{j}h_{ij}y_j+\eta_i.
\end{equation}
By definition:
\begin{equation}
\label{eq: app2} \gamma^{(0)}_j=v_j+\sum_{i=1, i\neq
j}^m(y_i-\hat{y}^{(0)}_i)b_{ij}.
\end{equation}
Now, replacing (\ref{eq: app1}) in (\ref{eq: app2}) results in:
\begin{equation}
\label{eq: app3} \gamma^{(0)}_j=v_j-\sum_{i=1, i\neq
j}^m\sum_{r}h_{ir}y_rb_{ij}+\sum_{i=1, i\neq
j}^my_ib_{ij}+\sum_{i=1, i\neq j}^m\eta_ib_{ij}.
\end{equation}
If we add and subtract the terms with $i=j$, then after some
simplifications and calculations, we have:
\begin{displaymath}
\gamma^{(0)}_j=v_j-\sum_{i=1}^mb_{ij}\sum_rh_{jr}y_r+\sum_rh_{jr}y_r+
\end{displaymath}
\begin{equation}
\sum_{i=1}^my_ib_{ij}-y_j+\sum_{i=1}^m\eta_ib_{ij}-\eta_j.
\end{equation}
It leads to the following matrix form:
\begin{equation}
\boldsymbol{\gamma}^{(0)}=\vb+(\Bb-\Ib)\boldsymbol{\eta}-(\Ib-\Bb+\Bb^T\Hb-\Hb)\yb.
\end{equation}
with $b_{ij}\triangleq<\boldsymbol{\phi}_i,\boldsymbol{\phi}_j>$
and $v_j\triangleq <\eb,\boldsymbol{\phi}_j>$. Using
$\vb=\Phib^T\eb$ and $\boldsymbol{\eta}=\Phib^\dagger\eb$, then we
have:
\begin{equation}
\label{eq: e2}
\boldsymbol{\gamma}^{(0)}=\boldsymbol{\Psi}\yb+\Lb\eb.
\end{equation}
where $\Lb=\Phib^T+(\Bb-\Ib)\Phib^\dagger$ and
$\boldsymbol{\Psi}=-\Bb^T\Hb+\Bb+\Hb-\Ib$. Using
$\Bb=\boldsymbol{\Phi}^T\boldsymbol{\Phi}$ and
$\Hb\triangleq\Phib^\dagger\Phib$, we have
$\Lb=2\Phib^T-\Phib^\dagger$ and $\boldsymbol{\Psi}=-\Ib+\Hb$.
Finally, (\ref{eq: e2}) results in (\ref{eq: initpar}).

\section{Sufficient condition for stability of Hard-BHTA}
\label{app2} For small values of $\sigma_{\gamma_j}$ in comparison
to $\sigma_r$, it can be seen that the threshold is proportional
to:
\begin{equation}
\mathrm{Th}_j\propto\sigma_{\gamma_j}\sqrt{\ln(\frac{p}{1-p}\frac{\sigma_r}{\sigma_{\gamma_j}})}.
\end{equation}
Therefore, if we define $x=\sigma_{\gamma_j}$ and
$c\triangleq\frac{p}{1-p}\sigma_r$, then we should investigate
monotonicity of the function $f(x)=x\sqrt{\ln(\frac{c}{x})}$.
Using the derivative of this function, it can be seen that the
function is decreasing for $x<\frac{c}{e}$. This means
$\sigma^2_{\gamma_j}<k^2\sigma^2_r$ where $k=\frac{p}{(1-p)e}$.
So, we should have:
\begin{equation}
\label{eq: suff1} \sigma^2_e+\sum_{i=1, i\neq
j}^mb^2_{ij}{\sigma^2}^{(n)}_{i,e_y}<k^2\sigma^2_r.
\end{equation}
Then, for the next iteration, it is obvious that the above
condition is satisfied because if $\alpha<1$ then:
\begin{equation}
\sigma^2_e+\alpha^2\sum_{i=1, i\neq
j}^mb^2_{ij}{\sigma^2}^{(n)}_{i,e_y}<k^2\sigma^2_r.
\end{equation}
We use the condition in (\ref{eq: suff1}) at initialization as the
sufficient condition for a decreasing threshold. Replacing the
initial variance of (\ref{eq: initpar}) in (\ref{eq: suff1}) for
$n=0$, then after some simple manipulations, leads to the
sufficient condition (\ref{eq: suff}).

\section*{ACKNOWLEDGEMENT}
We would like to thank the anonymous reviewers for their fruitfull
suggestions. Moreover, the first author would also like to thank
METISS group and INRIA/Rennes (IRISA) since most of this work was
done when the first author were there as a visiting researcher.

\end{document}